\documentclass{article}
\usepackage{color}
\usepackage{ulem}
\usepackage{siunitx}
\usepackage[english]{babel}
\usepackage{rotating}
\usepackage{longtable}
\usepackage{amsmath}
\usepackage{bm}
\usepackage{float}
\usepackage{amssymb}
\usepackage{graphicx}
\usepackage[colorinlistoftodos]{todonotes}
\usepackage[colorlinks=true, allcolors=blue]{hyperref}
\usepackage{comment}
\usepackage{natbib}
\usepackage{lineno}
\usepackage{sectsty}
\usepackage{setspace}

\usepackage{arxiv}

\usepackage[utf8]{inputenc} 
\usepackage[T1]{fontenc}    
\usepackage{hyperref}       
\usepackage{url}            
\usepackage{booktabs}       
\usepackage{amsfonts}       
\usepackage{nicefrac}       
\usepackage{microtype}      
\usepackage{lipsum}		
\usepackage{graphicx}
\usepackage{natbib}
\usepackage{doi}
\usepackage{color}
\definecolor{trackcolor}{rgb}{0,0,0}

\title{Systematic assessment of the effects of space averaging and time averaging on weather forecast skill}

\author{ {\hspace{1mm}Ying Li}\\
	Department of Mathematics\\
	University of Wisconsin-Madison\\
	Madison, WI, 53706, USA\\
	\texttt{li526@wisc.edu} \\
	\And
	{\hspace{1mm}Samuel N. Stechmann} \\
	Department of Mathematics \& Department of Atmospheric and Oceanic Sciences\\
	University of Wisconsin-Madison\\
	Madison, WI, 53706, USA\\
	\texttt{stechmann@wisc.edu} \\
}

\renewcommand{\headeright}{}

\begin{document}
\maketitle

\begin{abstract}
Intuitively, one would expect a more skillful forecast
if predicting weather averaged over one week instead of the weather averaged over one day, and similarly for different spatial averaging
areas.
However, there are few systematic studies of averaging and forecast skill
with modern forecasts, and it is therefore not clear how much improvement in forecast performance 
is produced via averaging.
Here we present a direct investigation of averaging effects, 
based on data from operational numerical weather forecasts.
Data is analyzed for precipitation and surface temperature, for lead times
of roughly 1 to 7 days, and for time- and space-averaging diameters of 
1 to 7 days and 100 to 4500 km, respectively.
For different geographic locations, 
the effects of time- or space-averaging can be different,
and while no clear geographical pattern is seen for precipitation,
a clear spatial pattern is seen for temperature.
For temperature, in general, time averaging is most effective
near coastlines, also effective over land,
and least effective over oceans.
Based on all locations globally, 
time averaging was less effective than
one might expect.
To help understand why time averaging may sometimes be
minimally effective, a stochastic model is analyzed as a
synthetic weather time series, and analytical formulas are
presented for the decorrelation time.
In effect, while time averaging creates a time series that is visually
smoother, it does not necessarily cause a substantial increase in the
predictability of the time series.
\end{abstract}


\section{Introduction}
\label{sec:intro}

It is natural to expect more accurate forecasts of
time-averaged quantities. For instance, one would expect that 
the rainfall averaged over one week can be forecast with greater skill
than the rainfall averaged over one day.
Likewise, one would expect that the rainfall averaged over a large
region (such as the area of a state or province) could be predicted
with greater skill than the rainfall over a smaller region
(such as a city or town). This conventional wisdom has a long history;
for instance, \citet{roads1986forecasts} writes,
``When meteorologists are asked to make a forecast 
for a month, a season, or even a couple of weeks in
advance, they usually quote from numerous predictability
studies that it is probably impossible to forecast 
instantaneous deviations from climatology this far into
the future. They might also point out that skillful
instantaneous forecasts are rarely issued by weather
services beyond a few days in advance. On the other hand,
they might suggest that it is possible to give a broad
overview of the upcoming weather by considering the
time-averaged behavior.''

The quote above is from 35 years ago, and it is important to note
that, since then, forecasting methods have undergone 
substantial changes. 
\begin{color}{trackcolor}
For instance, advances in data assimilation
methods have provided initial conditions that are more accurate,
and the models themselves are improved and used with 
higher resolution
\citep[e.g.,][]{btb15}.
If the initial condition or forecast is highly uncertain, then
the quote above seems to agree with basic intuition:
an instantaneous forecast may not be very accurate,
but its time-averaged behavior may be more skillful.
On the other hand, with a more accurate initial condition or forecast,
an instantaneous forecast could have substantially greater skill.
Consequently, these advances in
forecast methods raise the question,
How much improvement in predictability or
forecast performance is gained via time averaging,
in a modern forecasting setting?
\end{color}

Some past work, including some relatively modern work, 
has examined forecast skill for different spatial or temporal averaging
\citep[e.g.,][]{roads1986forecasts,r08ma,zwsh14},
although the studies were not set up in a way that allows direct
assessment of the conventional wisdom.
\begin{color}{trackcolor}
Other recent work, on the other hand, 
has been set up to allow a direct assessment
\citep{buizza2015forecast,van2020influence}.
Given that different setups have been used in different studies in
investigating spatial and temporal averaging,
it is worthwhile to clarify some important aspects of the setup.
\end{color}

\begin{figure}[h]
\includegraphics[height=0.36\textwidth,width=0.9\textwidth]{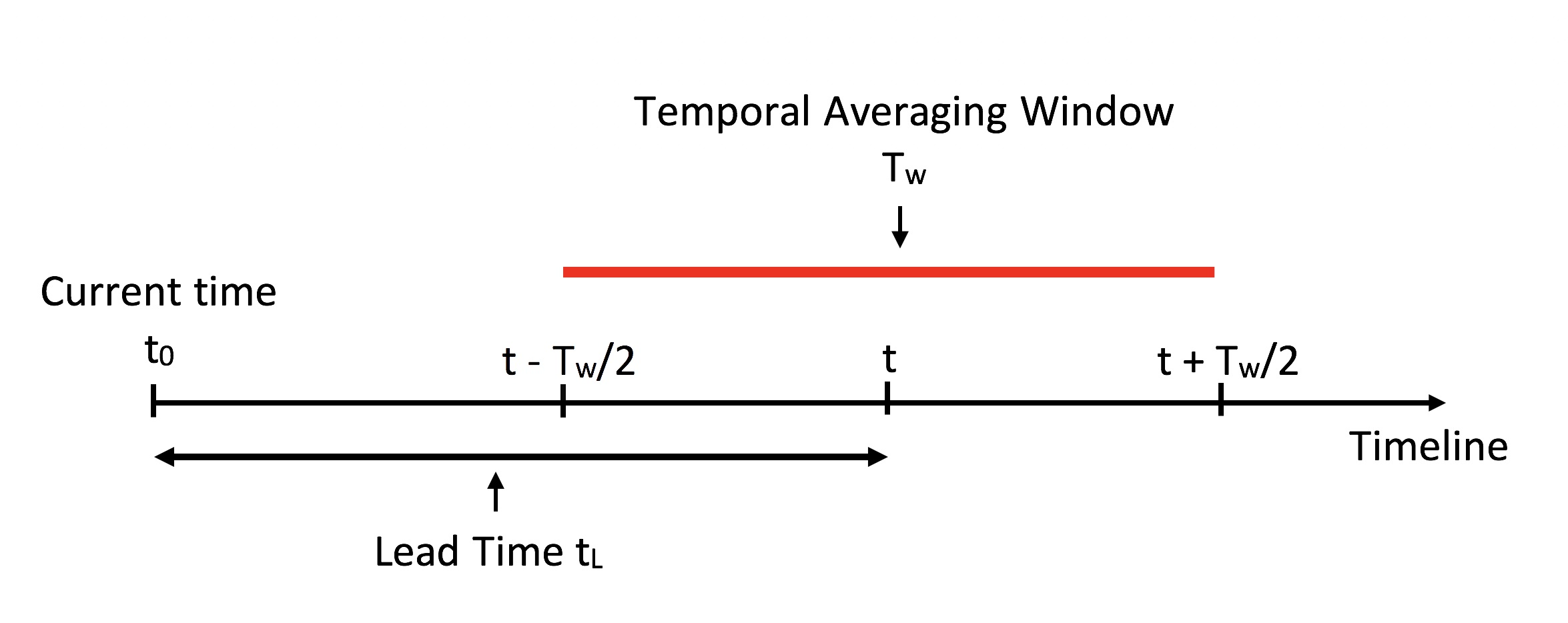}
\caption{Definition of the time-averaging window as a ``centered'' window, extending from $t-T_w/2$ to $t+T_w/2$. The horizontal axis represents forecast time from the initial condition at $t_0$.}
\label{timeavgdef}
\end{figure}

In particular,
to most clearly assess the conventional wisdom, 
it is necessary to compare forecasts in a way that holds all parameters fixed
(lead time, etc.) except for changes in the averaging window.
Furthermore, the time-averaging window should
extend in both directions around the target date,
as illustrated in Figure~\ref{timeavgdef}
\begin{color}{trackcolor}
(and discussed further in section~\ref{method}).
\end{color}
If the target date is time $t$, then a time average
should be performed from $t-(T_w/2)$ to $t+(T_w/2)$,
where $T_w$ is the duration of the time-averaging window.
If other conventions are used, they could introduce a bias into the
comparison. For instance, if the time average is performed
from $t$ to $t+T_w$, then the forecast skill will
decrease as $T_w$ increases, since the forecast is
performed further into the future. On the other hand,
\begin{color}{trackcolor}
a time average in the other direction, from $t-T_w$ to $t$,
is commonly used in numerical weather prediction (NWP)
to define an accumulation.
However,
\end{color}
if the time average is performed
from $t-T_w$ to $t$, then the forecast skill will
increase as $T_w$ increases, since the forecast is
performed closer to the current time.
To remove these biases and isolate the effects that are due to
only the time-averaging window duration $T_w$,
a time average should be performed over a ``centered'' window
from $t-(T_w/2)$ to $t+(T_w/2)$.
Such an assessment is the goal of the present paper.

\begin{color}{trackcolor}
A few recent studies have provided assessments of averaging
that use the ``centered'' window in Figure~\ref{timeavgdef}
\citep{buizza2015forecast,van2020influence},
and their setups have some similarities and some differences
in comparison to the present study.
\citet{buizza2015forecast}
use the centered (or midpoint) definition of the time-averaging window,
as in Figure~\ref{timeavgdef} here,
and they analyze data at all locations globally,
but they define spatial smoothing in a different way
(i.e., using spectral truncations), 
and they do not analyze precipitation.
\citet{van2020influence}
also use the centered (or midpoint) definition
of the time-averaging window,
and they use a spatial aggregation method based on
dissimilarity and hierarchical clustering.
They analyzed near-surface temperature 
(specifically, 2-m temperature), but not precipitation.
Also, they analyzed data over Europe, 
whereas the present study is global.
\end{color}

Besides using operational weather forecasts, 
one could also investigate the effects of time- and space-averaging
in other dynamical systems.
In some early work, simple stochastic models have been
used, and their statistics have been analyzed
after time averaging
\citep{munk1960smoothing,leith1973standard,jones1975estimating,roads1987estimate}.
While these early studies provide valuable information
about the statistics of time-averaged processes,
they do not provide a direct assessment of time-averaging
effects on forecasts as described above and in 
Figure~\ref{timeavgdef}.
In more recent work,
\citet{ls18}
made a direct assessment, as in Figure~\ref{timeavgdef}.
For simple stochastic dynamical systems, 
it was seen that spatial averaging did improve forecast performance,
whereas time averaging was somewhat limited at improving forecast performance, under the perspective of the simple stochastic setting. 

\begin{figure}[ht!]
\centering
\includegraphics[width=\textwidth]{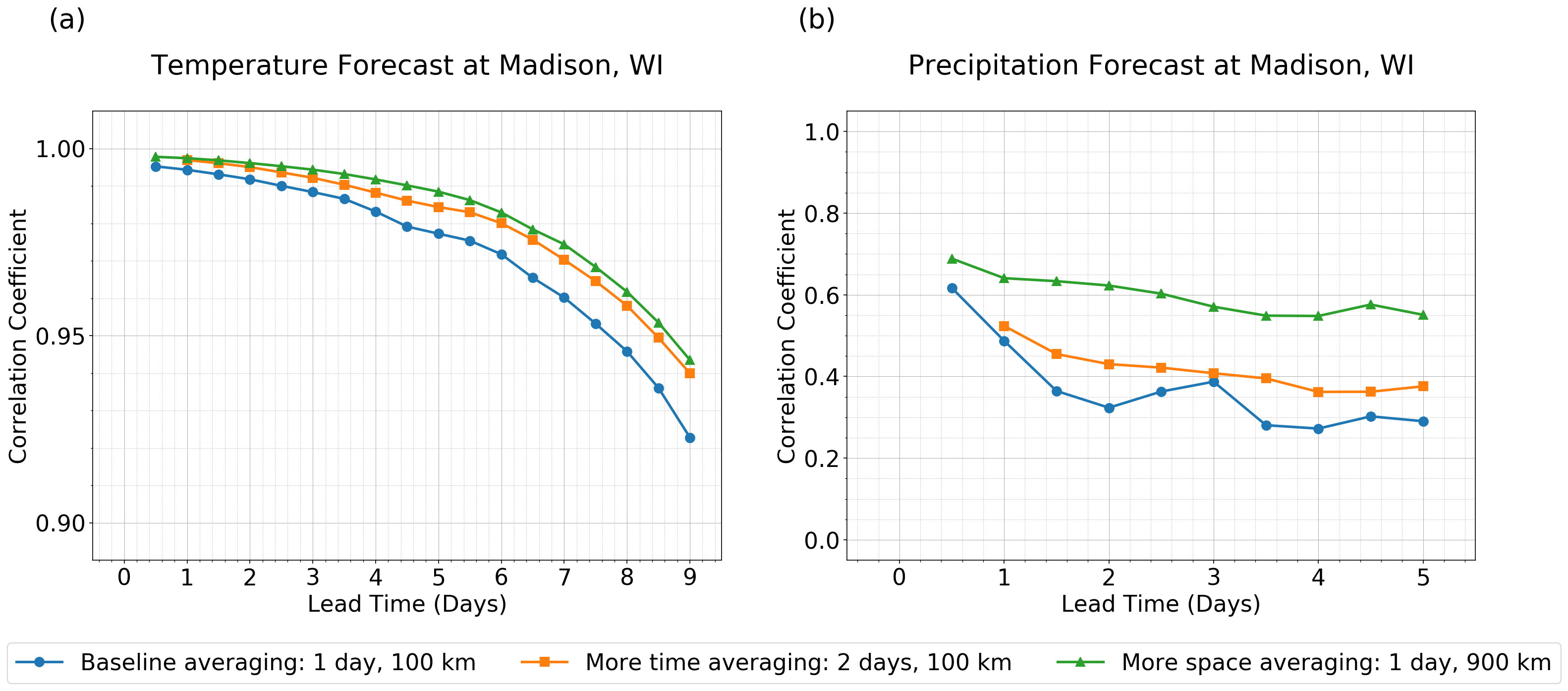}
\caption{Effects of time averaging and space averaging on forecast performance,
at the location of Madison, Wisconsin, USA, for (a) temperature forecasts
and (b) precipitation forecasts. For temperature, forecast performance is
improved by approximately the same amount for either the additional
time averaging or the additional space averaging. On the other hand,
for precipitation, the additional spatial averaging brings 
a much larger improvement in forecast performance than the additional 
time averaging.}
\label{fig:madison}
\end{figure}

As a preliminary look at numerical weather prediction data, 
Figure~\ref{fig:madison}
shows the effects of space- and time-averaging
at one location---our home city of Madison, Wisconsin.
While the methodology will be described in more detail below,
and while results may differ for different geographic locations,
Figure~\ref{fig:madison} offers a first glimpse
of the results.

Substantial differences can be seen in Figure~\ref{fig:madison}
for the effects of averaging.
A baseline case is shown in each panel to indicate a
somewhat minimal amount of averaging,
with a 1-day time-average to average over the diurnal cycle,
and with a 100-km spatial average that represents the
resolution of the data from the numerical forecast model.
Two additional curves are also shown in each panel:
one curve for forecasts of longer time averages of duration 2 days,
and one curve for forecasts of a larger spatial region of
diameter 900 km. 

In all cases, the additional averaging
leads to a enhanced forecast performance. However, for temperature,
the time- and space-averaging have similar impacts,
whereas for precipitation the time- and space-averaging
have substantially different impacts:
the spatial averaging causes a much greater improvement
in forecast performance. This is one illustration of how
the effects of time- and space-averaging can differ greatly
in different scenarios. 
\begin{color}{trackcolor}
The purpose of the present paper
is to investigate these effects systematically and in
greater detail, and to consider whether differences arise
due to other factors,
such as land versus ocean, tropics versus extratropics, etc.
\end{color}

The effects of time- and space-averaging could be investigated
in a number of different settings. Here, 
lead times of 0 to 9 days are investigated.
Such a choice is natural in the sense that this is the
classic forecasting problem, whereas longer lead times
are a more recent endeavor. 
(It would be interesting in the future
to carry out a similar study at longer lead times
such as subseasonal-to-seasonal prediction.)
Also, the data for these shorter
lead times (0 to 9 days) is available at higher spatial 
resolution than the data for extended-range and long-range
forecasts, and higher resolution will facilitate the study
of spatial averaging. 

Also, we note a distinction between two different approaches
to investigating forecast performance on different spatial scales.
As one approach, spectral decomposition is one common way to identify
different spatial scales, where each spatial scale can be
identified with a wavelength of a different spectral basis function,
such as a sinusoid or a spherical harmonic.
Spectral decompositions have long been used in analyzing
predictability on different scales
\citep[e.g.,][]{boer1984spectral,boer2003predictability,buizza2015forecast,toth2019weather}
and also have been used for various applications
such as to identify predictability
of different types of convectively coupled equatorial waves (CCEWs)
\citep{detal18,jetal18,ls20}.
As a second approach, one can use a regional spatial average
to identify a particular spatial scale, using an approach as in
Figure~\ref{timeavgdef} but for space instead of time.
One difference between these two approaches is that spectral
decompositions are global, whereas regional averages are local.
For the purposes of the present paper,
we would like to analyze forecast performance at different geographical 
locations, using time-averaging and/or space-averaging,
and for this purpose, the regional averaging technique will be
more suitable.

The rest of the paper is organized as follows.
The methods and data are described in
section~\ref{method}.
The impact of averaging on forecast performance is assessed in
section~\ref{sec:impact}, and in section~\ref{sec:theory}
a stochastic model is analyzed to provide some theoretical
insight.
A concluding discussion is in section~\ref{sec:concl}.
Additional sensitivity studies and theoretical calculations
are presented in Supporting Information.

\section{Methods}
\label{method}

In this section, we describe the data
(section~\ref{sec:data}) and the methods
(sections~\ref{sec:data-analysis-setup}--\ref{sec:comp-methods}),
including the methods for time- and space-averaging
(section~\ref{sec:data-analysis-setup}).

\subsection{Data}
\label{sec:data}

The operational numerical weather prediction data used here 
is from the Global Forecast System (GFS)
and for two variables: precipitation \begin{color}{trackcolor}{rate}\end{color} and surface temperature.
The 3 hourly and $1^\circ$ gridded global GFS dataset is used here, and it is chosen instead of the $0.5^\circ$ forecast dataset because the $1^\circ$ dataset is available out to 16-day lead times instead of 8-day lead times, thereby allowing assessment of longer available lead times and time-averaging windows.
The open source data can be downloaded at 
\url{https://www.ncdc.noaa.gov/data-access/model-data/model-datasets/global-forcast-system-gfs}. \textcolor{trackcolor}{Verification of the forecasts is done using GFS day-0 forecast data (also called the day-0 analysis) to serve as the truth for both precipitation rate and surface temperature for assessing the forecast performance. }
While observational data would be more desirable for the truth data for forecast verification,
many quantities are not available globally, and for this or other reasons
it is convenient and somewhat common to use the day-0 analysis or reanalysis as the ``truth'' for verification
\citep[e.g.,][]{roads1986forecasts,brankovic1990extended,simmons2002some,jetal18}.
For comparison, some additional tests will be reported with observational data of precipitation as the truth for verification.

For those additional tests, Global Precipitation Measurement (GPM) data are also used as a baseline truth, for comparison with the use of GFS day-0 analysis data as truth for precipitation. The 30-min research/final run version of GPM data with spatial resolution $0.1^\circ$ is used here for the variable ``precipitationCal'' (Multi-satellite precipitation estimate with gauge
calibration). The GPM data are fully available from $60^\circ$N - $60^\circ$S while much data is missing outside this range. Hence, only GPM data within $60^\circ$N - $60^\circ$S are used in the analysis here to ensure stable and reliable results. GPM data can be downloaded from \url{https://pmm.nasa.gov/data-access/downloads/gpm}.

As another type of sensitivity test, data has also been used for
anomalies from the seasonal cycle.
\begin{color}{trackcolor}
We have investigated three approaches to creating anomalies
from the seasonal cycle. In one approach to create the anomaly data,
\end{color}
a smoothed seasonal cycle has been subtracted from the GFS forecast data before moving forward with further analysis. For the seasonal cycle of precipitation rate, the enhanced monthly long term mean CPC Merged Analysis of Precipitation (CMAP) with $2.5^\circ$ latitude x $2.5^\circ$ longitude global grid is used (and it can be downloaded from \url{https://psl.noaa.gov/data/gridded/data.cmap.html}). For creating the seasonal cycle for surface temperature, GFS analysis data with lead 0 from January 1, 2005, to December 31, 2017 are used. In the preprocessing, the 6 hourly smoothed seasonal cycle is calculated via the annual mean and the first three harmonics at every location globally through discrete Fourier transform and inverse discrete Fourier transform. Also, a bilinear interpolation was used to interpolate the smoothed $2.5^\circ$ CMAP precipitation seasonal cycle to a $1^\circ$ grid to match the grids of the GFS dataset. After removing the smoothed seasonal cycle from the GFS forecast data, the remaining data represent anomalies from the seasonal cycle.
\begin{color}{trackcolor}
In this first approach, note that the seasonal cycle is defined as an
average over many years, and it could differ substantially from
the seasonal progression during the particular time period
considered here. For this reason, we also consider two other approaches
that approximate the seasonal progression in the particular
time period considered here.
In the second approach to create the anomaly data,
a 28-day average is applied to the time series in order
to define a low-frequency signal as an approximate seasonal progression.
The 28 days of averaging is split into 14 days of averaging 
in the past and 14 days of
averaging in the future. This averaged time series is then
subtracted from the original time series to define the
anomaly data. 
A third approach is also used, and it is identical to the
second approach, except a 14-day average is applied instead of
a 28-day average, and the 14-day average is an average
in the past 14 days, without any influence from future data.
The second approach may provide the best estimate of the low-frequency
or seasonal progression near each point in time,
although its use of future data may be less desirable for a
study of forecast performance. The third approach uses no future data,
although its estimate of the low-frequency or seasonal progression
may be somewhat skewed by its use of only past data (i.e., a
one-sided time window).
\end{color}

Four main assessments are conducted here,
and they cover a variety of quantities of interest (precipitation vs. surface temperature), datasets, time ranges (in line with the availability of the datasets), etc.:
    (1) 3.5 months of GFS precipitation forecast data from June 17, 2019 to October 08, 2019 are compared with GFS day-0 analysis data for precipitation. 
    (2) 11 months of GFS surface temperature forecast data from March 01, 2018 to January 28, 2019 are compared to GFS day-0 analysis data for surface temperature.
    \begin{color}{trackcolor}
    The durations of 3.5 months of precipitation data
    and 11 months of surface temperature data arise from the
    range of data that was available from the GFS website
    when we undertook our investigation.
    \end{color}
    (3) 11 months of GFS precipitation forecast data from November 01, 2018 to August 31, 2019 are compared with the GPM dataset.
    (4) The assessments 1 and 2 were also repeated with GFS precipitation/temperature data as anomalies from the seasonal cycle.
GFS forecast data for precipitation rate and surface temperature (Assessments 1 and 2) are mainly used in this main manuscript. Analysis involving the GPM dataset and the same analysis on the anomaly data of GFS (Assessments 3 and 4) are investigated as sensitivity tests. More details and results can be found in the Supporting Information. 

\subsection{Data Analysis Setup}
\label{sec:data-analysis-setup}

To describe the data analysis methods, the main specifications
are the choices of the time- and space-averaging windows,
and also some aspects of applying and comparing averages
between different types of datasets.

For calculating spatial averages, the averaging region is taken
to be the interior of a circle. (A square could also possibly be used, although
the orientation of a square would introduce some amount of
preferred directionality, whereas a circle is invariant under
rotations.) As the smallest circular region, we use a diameter
of 100 km, since it is roughly the same size as the 1$^\circ$ grid spacing 
of the GFS data.
For intuition, a 100-km-diameter circle has an area of
roughly 8,000 km$^2$, which is comparable to the area of the
state of Delaware (6,446 km$^2$).
For larger averaging areas, we consider diameters of 450 km, 900 km,
and so on, with an increment of 450 km, up to a diameter of
4,500 km. For intuition, a 4,500-km-diameter circle has an area of
roughly 64,000,000 km$^2$, which is
roughly the combined area of the five
largest countries (Russia, Canada, China, USA, and Brazil, together),
and is also approximately one-eighth
of the surface area of Earth (roughly 500,000,000 km$^2$).
Hence, a wide range of spatial averaging diameters is considered.

\begin{color}{trackcolor}
Note that applying spatial averages over very large areas
can sometimes lead to the concept of ``false skill''
\citep{hamill2006measuring,dorninger2018setup}.
When a very large averaging area is used, there could be
so many different weather systems contributing to the assessment
that it is difficult to ascertain the source of skill 
or lack thereof.
Therefore, care should be used in interpreting the results here
with large spatial averaging areas. 
In light of this,
most of the focus here will be on the smaller spatial averaging
areas, and the larger areas are included mainly as
a further exploration.
\end{color}

For time averaging, the averaging window is of duration $T_W$,
and the window is centered about the time of interest $t$.
Explicitly, the averaging window extends from $t-(T_w/2)$
to $t+(T_w/2)$, as illustrated in Figure~\ref{timeavgdef}.
As the smallest averaging window, we use a duration of
1 day, since it is somewhat common to assess forecast performance
of global models using daily-averaged fields
\citep[e.g.,][]{zwsh14,detal18}
and since diurnal variability is difficult to successfully
forecast in NWP models
\citep[e.g.,][]{berenguer2012diurnal}.
One difficulty of diurnal variability is that it is
often linked to details of convective
systems, such as convective initiation and propagation,
which occur on fine scales that are not well-represented on the 
grid of the GFS model, which has a horizontal grid spacing of
28 to 70 km.
For longer averaging durations, we consider window durations of
1.5 days, 2.0 days, and so on, with increments of 0.5 days,
up to a maximum of 6 days.
The 6-day maximum in averaging durations is chosen
in order to keep the entire averaging window in the future,
without extending back into the past, for a lead time of 3 days.
More precisely, for a 3-day lead time, a 6-day averaging window
would extend from day-0 to day-6, and any larger averaging
window would include data from the past.

\begin{color}{trackcolor}
Time averaging will produce an accumulated precipitation rate
over a longer period of time.
In the GFS forecast dataset, the precipitation data used is the precipitation rate per 3 hours.  For the shortest time-averaging window of 1 day, 8 consecutive precipitation rate data points (one data point every 3 hours) are added together to form the total precipitation over 24-hours in our paper. The different units of precipitation per 3 hours or per 1 day, etc., do not have an influence on the the pattern correlation that is used as the forecast performance metric, since the pattern correlation definition includes a rescaling by the standard deviation of the data. For other time-averaging windows, such as a time-averaging window of 2 days, the precipitation data that is analyzed is the precipitation accumulated over those 2 days within the time-averaging window.
\end{color}

Different combinations of lead times and averaging windows
will be investigated.
For precipitation, forecast performance degrades substantially
for lead times of roughly 0 to 5 or 6 days,
so the main lead times of interest will be 0 to 6 days.
For surface temperature, lead times of 0 to 9 days
will be investigated, since surface temperature forecasts 
remain skillful
(e.g., have correlation coefficient of 0.5 or greater)
for lead times of roughly one week or longer, depending on 
geographical location. Also, the GFS dataset used here
provides data out to maximum lead times of roughly two weeks,
which is long enough to allow asessments with lead times of 9 days
and time-averaging windows of 6 days.
\begin{color}{trackcolor}
For each lead time, many different time-averaging windows are
considered. For example, for a lead time of 9 days, 
time-averaging windows are considered of duration
1 day, 2 days, 3 days, 4 days, 5 days, and 6 days.
\end{color}

To achieve the comparability between the forecast dataset and the truth data and a well-defined setup for the time averaging window, some data processing is implemented before calculating the forecast performance with different averaging windows.  

First, for comparisons involving spatial averaging window larger than or equal to 900km, 30 min GPM data are averaged at each latitude and longitude to be coarsened from $0.1^\circ$ to $1^\circ$ at the very beginning. The GFS forecast/analysis data and GPM data are then averaged over a circular region with diameter of 450 km (or 900 km, ... , 4500 km) followed by the comparison.

Second, for comparisons involving time averaging, the baseline is with spatial averaging with a diameter of 100 km. For high resolution GPM data with $0.1^\circ$, the data is directly averaged over a circular region with diameter of 100km to create a preprocessed GPM data with 100km spatial averaging with the guarantee of enough sample points within 100km for each integer latitude and longitude. For GFS data with $1 ^\circ$, an underlying problem arising for those tropical locations is that they don't have enough sample points within 100 km spatial averaging due to the coarse grid. To achieve a better estimate, a bilinear interpolation has been applied to the GFS dataset to create new data with $0.2^\circ$ grids, which is then  spatial averaged over a circular region with diameter of 100km as a new preprocessed GFS dataset. Using the preprocessed datasets, we can move forward to 
do time averaging with 1 day, 1.5 days, ..., 6 days.

\begin{color}{trackcolor}{In comparing the analysis signal and the forecast signal, the same averaging is applied to both the analysis data and the forecast data.}
\end{color}

\subsection{Evaluating Forecast Performance}

The Pearson correlation coefficient $\rho$ is used here as the main metric for evaluating the forecast performance. The coefficient $\rho$ can be expressed as 
\begin{equation}
\rho(X(t),Y(t))=\frac{cov(X(t),Y(t))}{\sigma_X\sigma _Y},
\end{equation}
where $cov(X(t),Y(t))=\mathbb{E}[(X(t)-\mathbb{E}[X(t)])(Y(t)-\mathbb{E}[Y(t)])]$ is the covariance between $X(t), Y(t)$ and $\sigma_X,\sigma_Y$ are the standard deviations of $X,Y$ respectively. In practice, the expected value $\mathbb{E}$ is calculated empirically as a time average, so $\rho(X(t),Y(t))$ is  a constant that doesn't change over time. In particular, in a situation of analyzing the real data, we have the true data from the true signal as $X_1,X_2,\cdots,X_N$ and our corresponding predictions $Y_1,Y_2,\cdots,Y_N$, each given at $N$ points in time. The correlation coefficient is calculated by
\begin{equation}
\rho=\frac{\frac{1}{N}\sum_{i=1}^N(X_i-\frac{1}{N}\sum_{j=1}^N X_j)(Y_i-\frac{1}{N}\sum_{j=1}^N Y_j)}{(\frac{1}{N}\sum_{i=1}^N(X_i-\frac{1}{N}\sum_{j=1}^N X_j)^2)^{1/2}(\frac{1}{N}\sum_{i=1}^N(Y_i-\frac{1}{N}\sum_{j=1}^N Y_j)^2)^{1/2}},
\end{equation}

\begin{color}{trackcolor}
Other measures of forecast performance,
such as root-mean-square error and 
continuous ranked probability score, have also been used
in other studies
\citep{buizza2015forecast,van2020influence}.
Note that the pattern correlation will measure the correlation
of the time series patterns, but it will not measure intensity
and bias. The pattern correlation is used here for the
simplicity of its definition. In a similar study, but with
idealized stochastic models rather than numerical weather 
prediction model data, the conclusions about
time- and space-averaging effects were essentially the same
when measured based on pattern correlation or a scaled version
of root-mean-square error
\citep{ls18}.
Also note that skill scores are commonly defined as metrics
that measure performance relative to a reference or benchmark
\citep[e.g.,][]{wilks2011statistical,jolliffe2012forecast}.
While pattern correlation is a ratio of a covariance, cov$(X(t),Y(t))$,
and the climatological reference standard deviations, 
$\sigma_X \sigma_Y$, it is common for skill scores
to be defined in other ways. When the term ``skill'' is used
in what follows, it is meant as a more brief terminology in place
of ``pattern correlation,'' and will frequently be noted as such
in those cases.
\end{color}

\subsection{Computational Methods}
\label{sec:comp-methods}

A substantial computational cost is required in order to calculate
forecast performance with various spatial and time averaging windows and at all global locations. 
For example, consider the assessment of the forecast performance for GFS temperature data with different time averaging windows (1 day to 9 days) and different spatial averaging, using as input a global dataset with approximately one year of data at 6 hourly time intervals. Furthermore, the GFS temperature forecast dataset includes lead times from lead 0 to lead 16 days. To analyze such a dataset, memory of about 70-75 GB is needed for each latitude to keep the running time of the calculation within about 8 hours for one latitude. 

To manage these costly calculations, high-throughput computing (HTC) is used
\citep{condor-practice}.
HTC is designed for a different type of computing task
in comparison to high-performance computing (HPC).
While HPC involves a
coupled computing environment with frequent communication
between processors, HTC is designed for computing tasks
that involve minimal communication.
Since the global data here can be partitioned into
several independent regions (e.g., for different latitudes),
HTC is a useful computing paradigm for simultaneously analyzing 
data from each region independently.
Our calculations were performed using the resources of the
Center for High Throughput Computing (CHTC)
at the University of Wisconsin--Madison.
By using HTC, we are able to
evaluate forecast performance at all latitudes simultaneously, 
to reduce the overall calculation time. 

\section{Impact of averaging on forecast performance}
\label{sec:impact}

The earlier plots in Figure~\ref{fig:madison}
used a particular choice of the averaging window sizes
and also a particular location of Madison, Wisconsin.
To investigate further, in this section, an expanded analysis
is carried out to consider whether the results are
similar or different for different averaging window sizes
and different locations.

\begin{figure}[ht!]
\centering
\includegraphics[width=\textwidth]{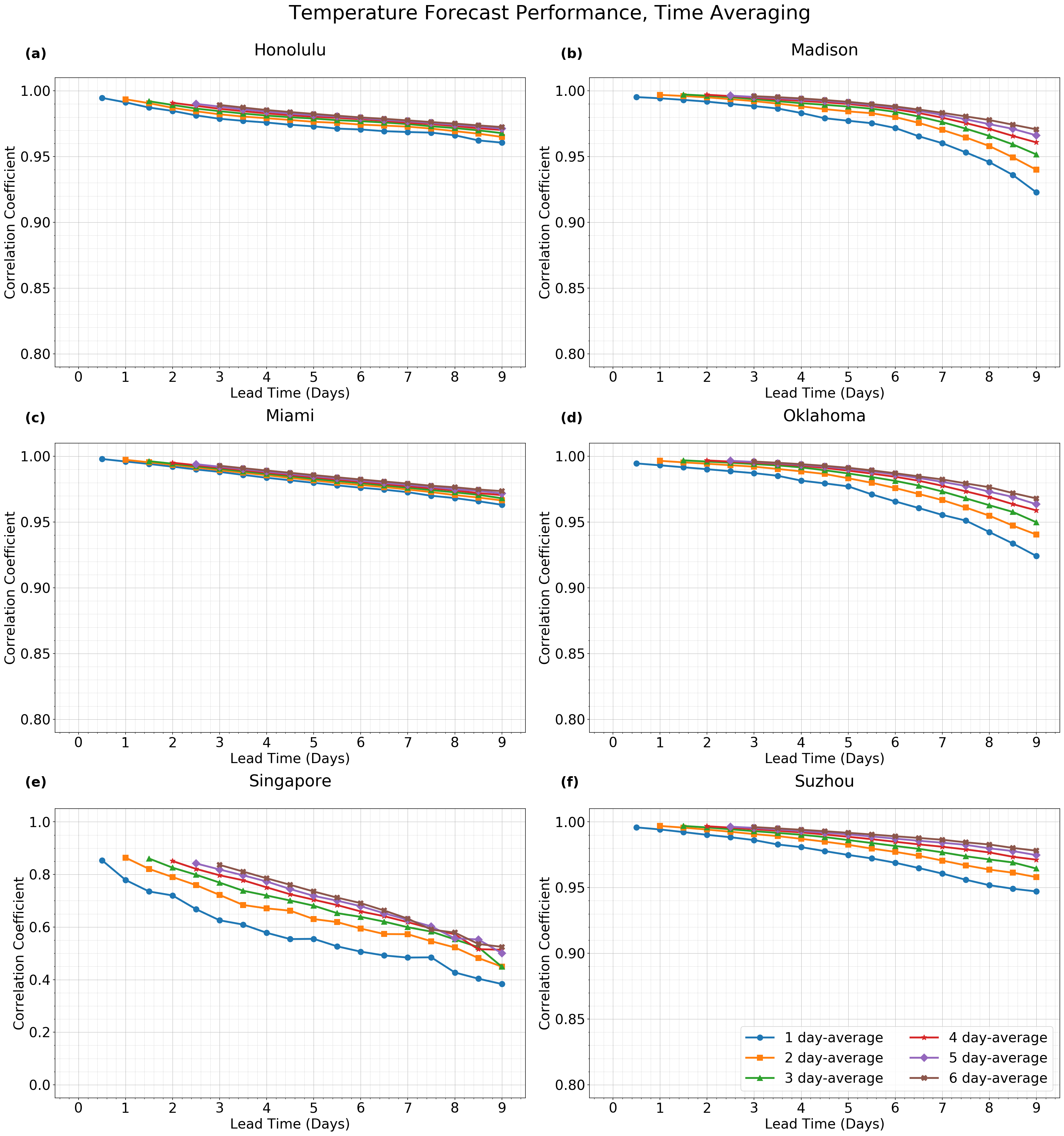}
\caption{Effects of time averaging on forecast performance, for surface temperature, 
\begin{color}{trackcolor}{with 100-km spatial averaging,}\end{color}
at several locations around the globe:
(a) Honolulu, Hawaii, (b) Madison, Wisconsin, (c) Miami, Florida,
(d) Oklahoma City, Oklahoma, (e) Singapore, and (f) Suzhou, China.
Compare to the effects of spatial averaging from Figure~\ref{fig:temp-space-avg}.
}
\label{fig:temp-time-avg}
\end{figure}

The effect of time averaging on surface temperature forecasts
is shown in Figure~\ref{fig:temp-time-avg}.
In order to assess results for a variety of different
geographical and climatic conditions
(e.g., land vs. sea, tropical vs. extratropical, etc.),
data is shown from six different locations:
Honolulu, Hawaii; Madison, Wisconsin;
Miami, Florida; 
Oklahoma City, Oklahoma;
Singapore; and Suzhou, China.
\begin{color}{trackcolor}
Six different curves are plotted for each location,
and each of the six curves is a different color, and
each curve
represents the forcast performance under
one of six cases of the time-averaging window duration:
1 day, 2 days, 3 days, 4 days, 5 days, or 6 days.
By comparing the different curves at a single location,
one can analyze how much improvement arises in the 
forecast performance as the duration of the time-averaging
window is increased.
\end{color}
From this figure, one can see that time averaging can have
substantially different effects, depending on the
particular geographic location.
For instance, relatively little improvement in forecast performance
is seen for Honolulu and Miami, whereas a much greater
improvement is seen for the other locations.

\begin{figure}[ht!]
\centering
\includegraphics[width=\textwidth]{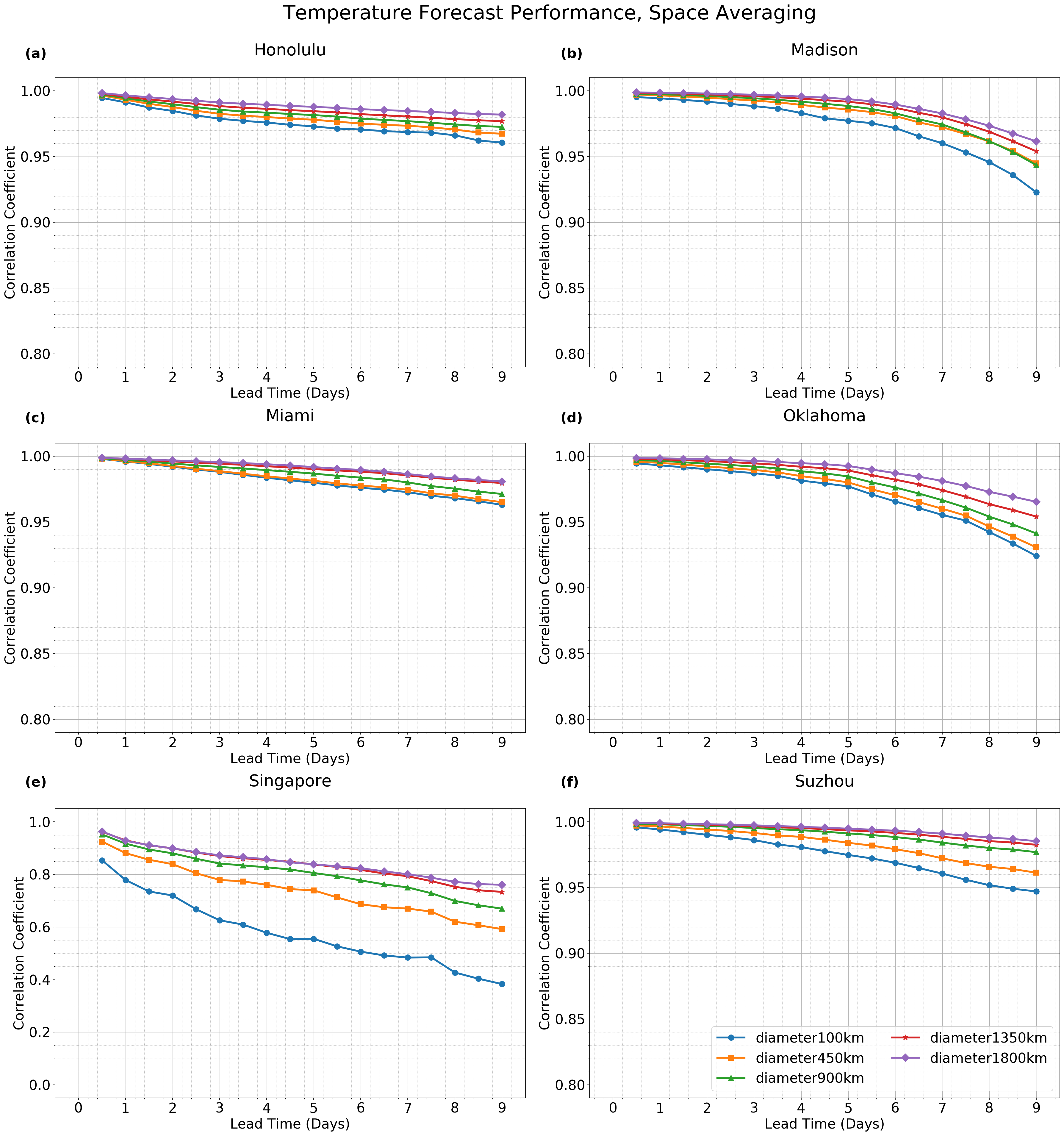}
\caption{Effects of spatial averaging on forecast performance, for surface temperature, 
\begin{color}{trackcolor}{with 1-day time averaging,}\end{color}
at several locations around the globe:
(a) Honolulu, Hawaii, (b) Madison, Wisconsin, (c) Miami, Florida,
(d) Oklahoma City, Oklahoma, (e) Singapore, and (f) Suzhou, China.
Compare to the effects of time averaging from Figure~\ref{fig:temp-time-avg}.}
\label{fig:temp-space-avg}
\end{figure}

For space averaging (Figure~\ref{fig:temp-space-avg}),
similar statements are true, although there is less of a
distinction in forecast improvement between
Honolulu and Miami on the one hand and the rest of the locations
on the other hand. 

\textcolor{trackcolor}{Note, for aiding the comparison, that the baseline case (blue curve) in Figure~\ref{fig:temp-space-avg} is the same as in Figure~\ref{fig:temp-time-avg}, as both use 1-day time averaging and 100-km spatial averaging.}

\begin{color}{trackcolor}
Also note that, for the time averaging results in
Figures~\ref{fig:temp-time-avg} and \ref{fig:temp-space-avg},
the forecast performance for lead time $L$ is only reported if 
the lead time is
greater than the half-width of the time-averaging window
(i.e., if $L>=T_w/2$). This condition will restrict attention to scenarios
where all the forecast data lies in the future,
so that it is a true ``forecast'' in the usual sense of the word.
For instance, for a lead time of 1 day and a time averaging window
of $T_w=4$ days, the time averaging window would include 
data from 3 days into the future and 1 day into the past,
so no data is plotted in this case.
\end{color}

\begin{figure}[ht!]
\centering
\includegraphics[width=\textwidth]{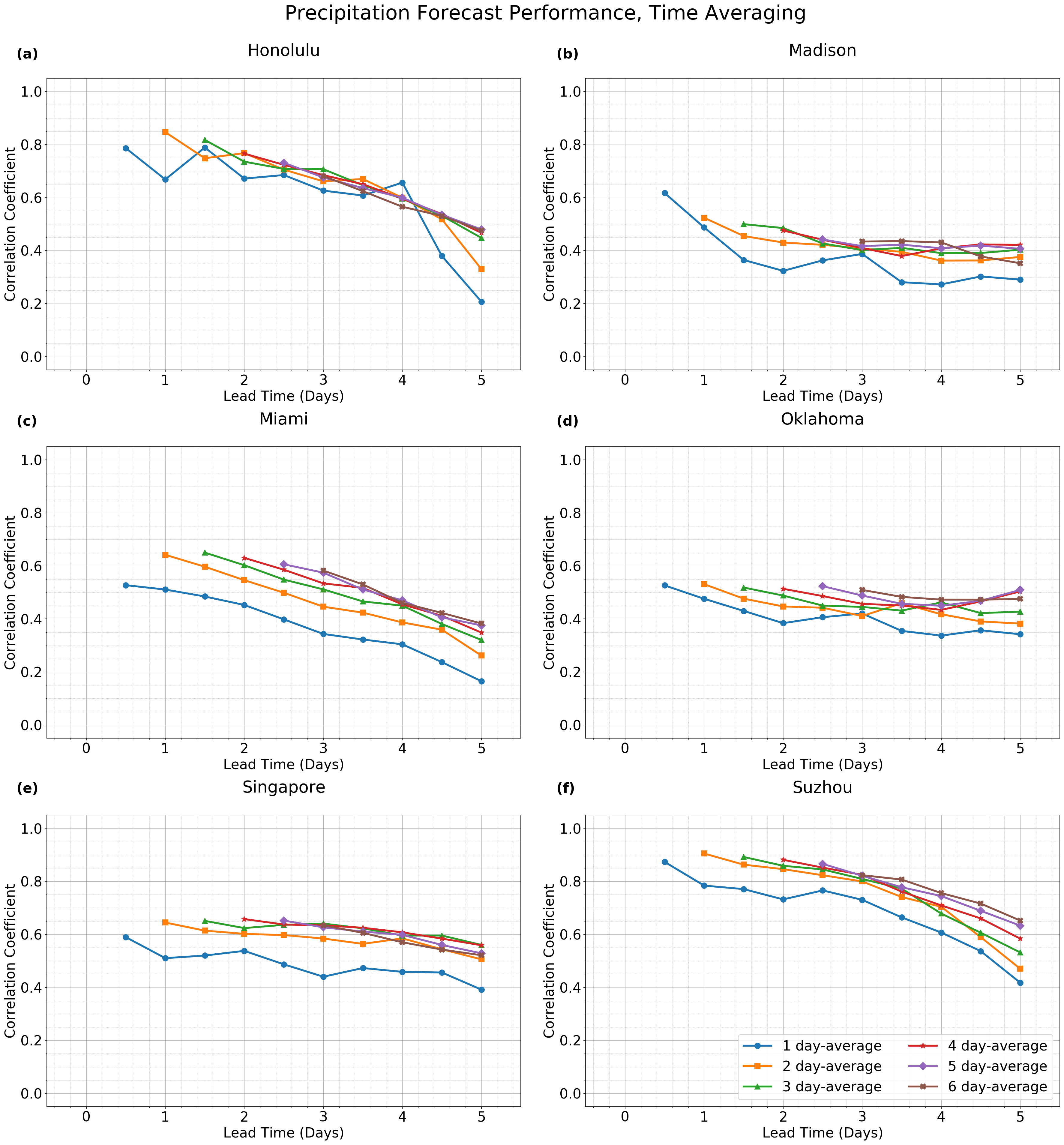}
\caption{Effects of time averaging on forecast performance, for precipitation, \textcolor{trackcolor}{with 100-km spatial averaging,}
at several locations around the globe:
(a) Honolulu, Hawaii, (b) Madison, Wisconsin, (c) Miami, Florida,
(d) Oklahoma City, Oklahoma, (e) Singapore, and (f) Suzhou, China.
Compare to the effects of spatial averaging from Figure~\ref{fig:precip-space-avg}.}
\label{fig:precip-time-avg}
\end{figure}

\begin{figure}[ht!]
\centering
\includegraphics[width=\textwidth]{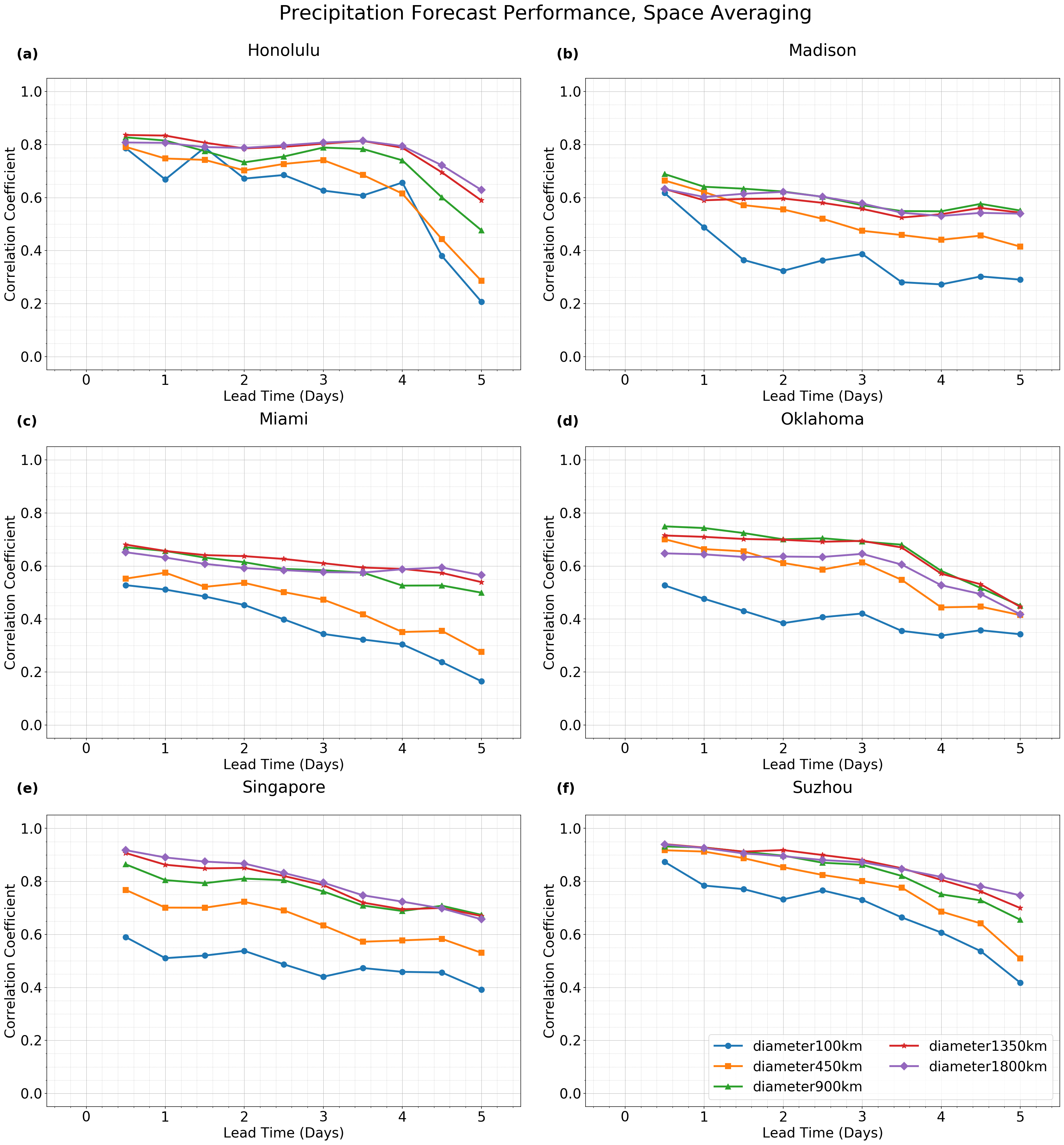}
\caption{Effects of spatial averaging on forecast performance, for precipitation, \textcolor{trackcolor}{with 1-day temporal averaging},
at several locations around the globe:
(a) Honolulu, Hawaii, (b) Madison, Wisconsin, (c) Miami, Florida,
(d) Oklahoma City, Oklahoma, (e) Singapore, and (f) Suzhou, China.
Compare to the effects of time averaging from Figure~\ref{fig:precip-time-avg}.}
\label{fig:precip-space-avg}
\end{figure}

For precipitation, similar plots are shown in
Figure~\ref{fig:precip-time-avg} for time averaging
and in Figure~\ref{fig:precip-space-avg} for spatial averaging.
In these plots,
time averaging improves forecast performance the least at Honolulu
and the most at Miami, Singapore, and Suzhou. 
Space averaging also improves performance the least at Honolulu,
along with Suzhou,
and the most at Singapore. 
These statements are based on the general spread of the curves
over all time- and space-averaging windows that are considered.
Some details of these statements could change depending on
the type of comparison. For instance, if a more restricted comparison
is made between only two space-averaging diameters of 100 km and
450 km, then the greatest improvement in forecast performance is seen at Oklahoma City
and a great (but slightly less) improvement is seen at Singapore.
In any case, these plots illustrate that time- and space-averaging
can have different impacts at different locations.

\clearpage

\begin{figure}[ht!]
\centering
\includegraphics[width=\textwidth]{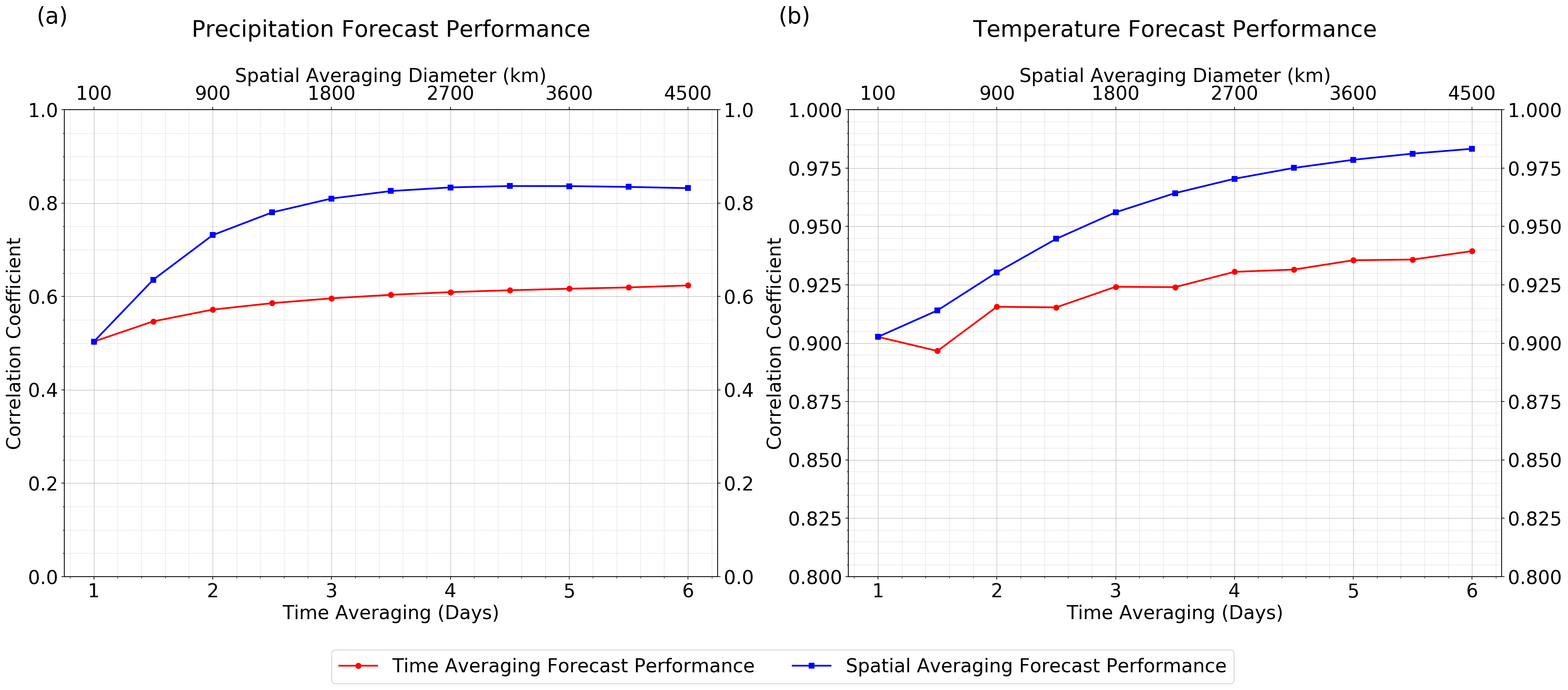}
\caption{Effect of time and space averaging on forecast performance, averaged globally. \textbf{(a)} Precipitation, at a lead time of 3 days. \textbf{(b)} Surface temperature, at a lead time of 7 days.}
\label{fig:global-avg}
\end{figure}

While Figures~\ref{fig:temp-time-avg}--\ref{fig:precip-space-avg}
showed geographical differences, the globally averaged behavior
is summarized in Figure~\ref{fig:global-avg}.
To simplify the plots, only one lead time is shown for each variable:
3 days for precipitation, and 7 days for temperature.
The forecast performance is reported as a correlation coefficient,
although it is a global average of correlation coefficients.
In other words, the skill (correlation coefficient) 
was first calculated for each location on the Earth,
and then the skill was averaged over all locations globally, 
leading to the the globally averaged forecast skill that is plotted in Figure~\ref{fig:global-avg}. 

In Figure~\ref{fig:global-avg},
for precipitation, a substantial improvement in forecast performance
is obtained from spatial averaging. In particular,
the skill (correlation coefficient) increases from
roughly 0.5 to roughly 0.8 upon increasing the spatial-averaging
diameter from 100 km to 1350 km. 
Large improvements in forecast performance are also seen in temperature
due to spatial averaging.
(Note the different axis range for temperature, for which
the correlation coefficient is always larger and above 0.9,
since temperature is more predictable than precipitation.)
On the other hand, for time-averaging, the correlation coefficient
increases, but by a lesser amount. For precipitation, for instance,
the increase in skill (correlation coefficient) is from 0.5 to roughly 0.6 
upon increasing the time-averaging window to nearly one week
(6 days).  To achieve such a skill level with spatial averaging,
a space-averaging diameter of only 450 km is needed.

\begin{color}{trackcolor}
Note that precipitation analysis data can be less accurate
at high latitudes, and Figure~\ref{fig:global-avg}
was created using precipitation analysis data over the
entire globe. As a sensitivity study and comparison,
GPM precipitation data has also been used for the 
forecast verification, and the data is restricted 
in latitude from only 60S to 60N. The result of this
sensitivity study is presented in the Supporting Information,
and the main features are similar to Figure~\ref{fig:global-avg}a.

It is interesting to note that 
\citet{buizza2015forecast}
show, in their Figure~8, only minimal increase in the
forecast skill horizon from increasing amounts of
spatial filtering. On the other hand, Figure~\ref{fig:global-avg}
here displays substantial improvement of forecast 
performance from increasing amounts of spatial averaging.
The difference between the results of \citet{buizza2015forecast}
and the results here are possibly related to the different
methods of spatial filtering. \citet{buizza2015forecast}
use a spectral truncation as a spatial filter, whereas the
present paper uses a local, regional spatial average.
Other factors could also contribute to the differences,
such as the different physical fields that are analyzed.
\end{color}

In the earlier plots, results were shown for several locations in
Figures~\ref{fig:temp-time-avg}--\ref{fig:precip-space-avg}
and for a global average in Figure~\ref{fig:global-avg}.
As another perspective, it would be interesting to view a map
of every location on Earth. To aide in the creation of the map,
the time-averaging and space-averaging results will be combined
together and contrasted by using a ratio $r$ that is defined as
$r = \Delta_x \rho/\Delta_t \rho$,
where $\rho$ is the forecast skill (correlation coefficient),
and $\Delta_t \rho$ is the change in forecast performance
due to an increase in time-averaging window duration
(from 1 day to 2 days),
and $\Delta_x \rho$ is the change in forecast performance
due to an increase in spatial-averaging window diameter
(from 100 km to 900 km).
\begin{color}{trackcolor}
We choose averaging windows of 1 day and 100 km because they
are the smallest window sizes considered here, and we
consider increases to 2 days and 900 km because they 
are somewhat small increases but still large enough to 
show changes in forecast performance,
as seen in Figures~\ref{fig:temp-time-avg}--\ref{fig:precip-space-avg}.
Also, the ratio of 800 km versus 1 day is approximately 10 m s$^{-1}$,
which is a typical value of a characteristic velocity of
synoptic scale or mesoscale fluctuations.
\end{color}
Hence, in this setup, if $r>1$, 
then spatial averaging is more effective than
time averaging at improving forecast performance, whereas if $0<r<1$,
then time averaging is more effective than spatial averaging
at improving forecast performance, at least for the selected averaging
windows of 1 to 2 days and 100 to 900 km. 

\begin{figure}[ht!]
\centering
\begin{minipage}{\textwidth}
\textbf{\textbf{(a)}}\\
\includegraphics[height=0.45\textwidth,width=\textwidth]{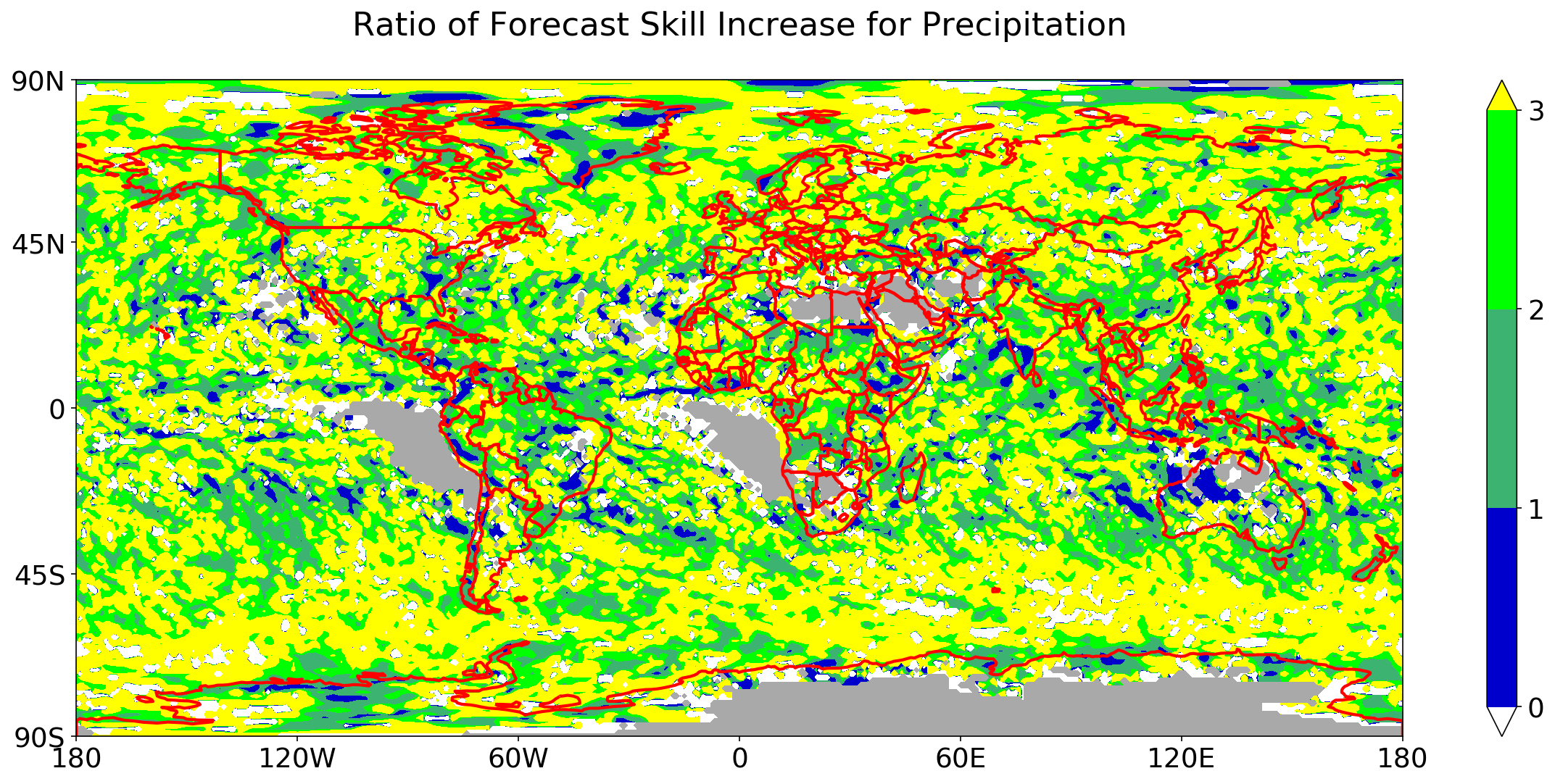}
\end{minipage}
\begin{minipage}{\textwidth}
\textbf{\textbf{(b)}}\\
\includegraphics[height=0.45\textwidth,width=\textwidth]{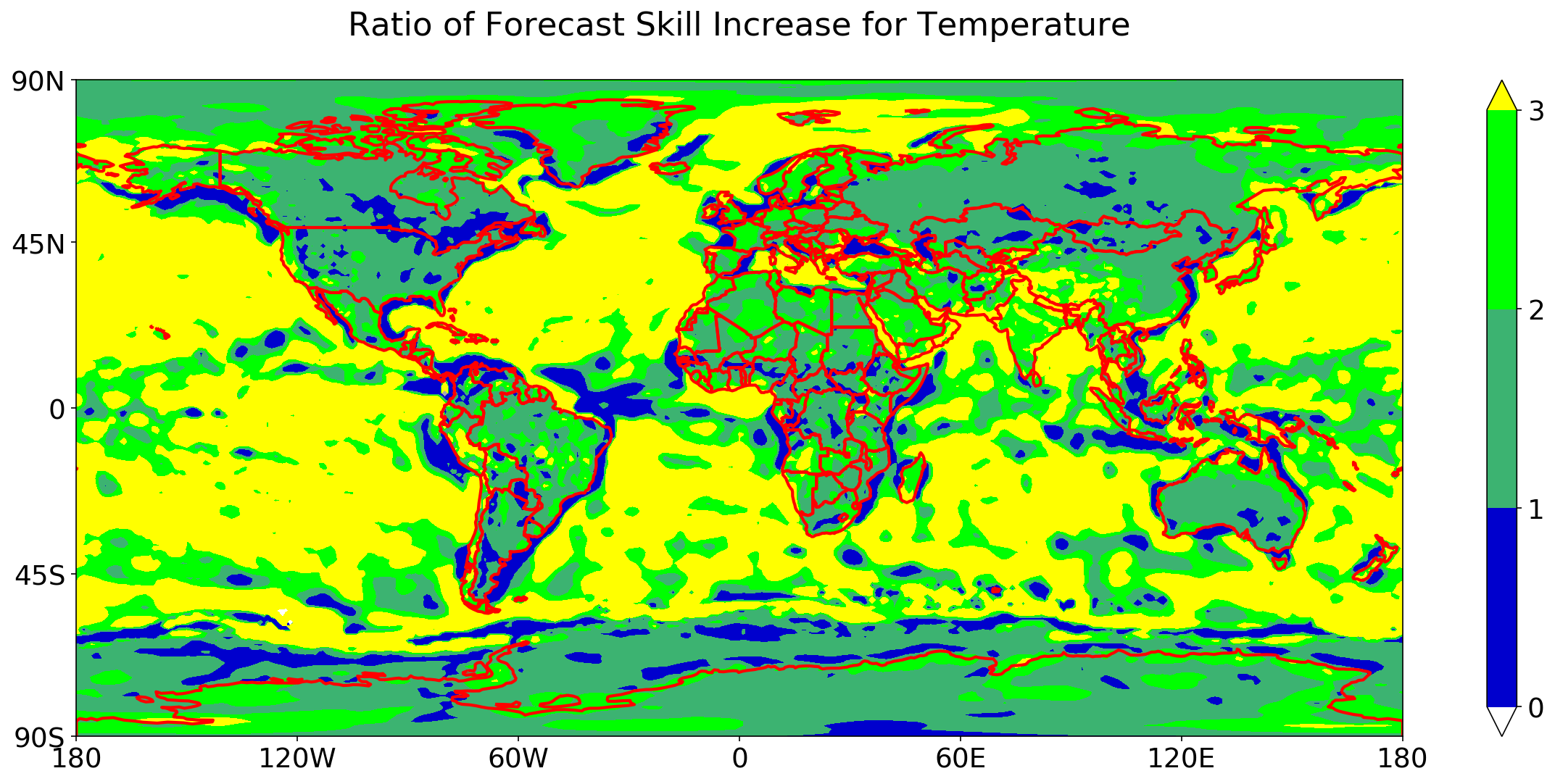}
\end{minipage}
\caption{Global map of the ratio of forecast skill increase,
defined as $r=\Delta_x \rho/\Delta_t \rho$,
where $\Delta_x\rho$ is the change in forecast skill (pattern correlation) due to 
increased spatial averaging, and
$\Delta_t\rho$ is the change in forecast skill (pattern correlation) for
increased time averaging. 
\textbf{(a)} Precipitation. \textbf{(b)} Surface temperature. Detailed explanations for the white and grey colors can be found in the text.}
\label{fig:maps}
\end{figure}

Figure~\ref{fig:maps} shows that, at nearly all locations worldwide,
$r>1$ and therefore spatial averaging is more effective
than time averaging at improving forecast performance,
at least for the selected averaging
windows of 1 to 2 days and 100 to 900 km.
In further detail, space averaging is, in fact, much more effective,
since $r>2$ or 3 over much of the map.

Note that some manual color settings are used for some special or extreme cases in creating Figure~\ref{fig:maps} to better present the comparison between time averaging and spatial averaging. Areas are marked with white if a negative ratio $r$ occurs; such a situation corresponds to improved forecast performance by spatial averaging but the degraded forecast performance by time averaging. In the opposite scenario, when a forecast increase is seen for time averaging and a  forecast performance decrease for spatial averaging, the ratio $r$ is manually set to be 0.001 for visualizations with blue colors to represent better forecast improvement through time averaging. Grey areas are locations where either missing values are present or where the forecast performance degrades for both spatial and time averaging, since this latter scenario should be distinguished from positive $r$ values due to forecast performance improvements for both spatial and time averaging. To summarize, in the global map presented in the paper, only blue color represents a better forecast improvement for the time averaging, grey color represents locations where a comparison is invalid due to missing values or that forecast performance degrades due to time averaging and also degrades due to spatial averaging, and all the other colors (yellow, forest green, medium sea green, white colors) are locations where spatial averaging has increases in forecast performance compared to the time averaging.

Some interesting geographical features can also be seen in Figure~\ref{fig:maps}.
For instance, for precipitation, in Figure~\ref{fig:maps}a, the
data has a random-looking geographical pattern.
Over much of the map for precipitation,
spatial averaging is three times more effective ($r>3$, yellow areas) than time averaging at improving the forecast performance. It is only in very few small regions that time averaging performs more efficiently (blue areas). 
\begin{color}{trackcolor}
Also, perhaps the main geographical features in the map for
precipitation are the grey regions, which are colored grey
due to limited precipitation over, for instance, the
Sahara desert, the Arabian desert, and the stratocumulus
cloud decks off the western coasts of Africa and South America.
\end{color}
For surface temperature, in Figure~\ref{fig:maps}b, an interesting characteristic is an apparent land--sea contrast:
time averaging is least effective
compared to spatial averaging ($r>3$, yellow color) over ocean regions,
less effective ($1<r<3$, green colors) over land regions,
and more effective ($0<r<1$) over coastal regions.
These geographic features are so apparent in the spatial map
of $r$ values that the outlines of the 
continents arise naturally in Figure~\ref{fig:maps}b.
Time averaging is more effective than spatial averaging
at only a very small number of locations, along the coastlines
of the continents; this is possibly due to the unique conditions
associated with coastal microclimates.

\begin{color}{trackcolor}
Additional sensitivity studies have also been carried out
to analyze anomalies from a seasonal cycle.
The seasonal cycle was defined using three different methods,
as described in section~\ref{method}.
One reason to analyze anomalies from the seasonal cycle
is that one may want to know the skill in predicting
only the day-to-day fluctuations, whereas the raw data 
includes seasonal fluctuations that will also influence
the pattern correlation in the forecast verification.
The results of the sensitivity studies are shown in
plots in the Supporting Information. 
One difference that arises for anomalies from the seasonal cycle
is that the forecast skill (pattern correlation) is lower
for anomalies. Another difference is that the geographical
pattern and land--sea contrast in Figure~\ref{fig:maps} 
could be more pronounced or less pronounced, depending on the 
method of defining the seasonal component.
Further study would be helpful to investigate the 
land--sea contrast in more detail.
One similarity across different tests is the 
greater effectiveness of spatial averaging compared to
time averaging at increasing the forecast skill (pattern correlation),
which is seen in global averages in Figure~\ref{fig:global-avg}
and also in all of the sensitivity studies.
\end{color}

\section{Theoretical investigation}
\label{sec:theory}

In the results in Figures~\ref{fig:madison}--\ref{fig:maps},
it was seen that time averaging was not as effective as space averaging
at improving forecast performance, at least over the averaging windows
considered here on meso-to-synoptic scales from 1 to 6 days and
100 to several thousand kilometers.
From basic intuition, one would expect that time averaging should provide a
smoother time series, and a smoother time series should be
more predictable; however, such enhanced predictability was more
limited than one might expect.

\begin{figure}[!hbtp]
\centering
\begin{minipage}{0.45\textwidth}
\textbf{\textbf{(a)}}\\
\includegraphics[height=0.9\textwidth,width=\textwidth]{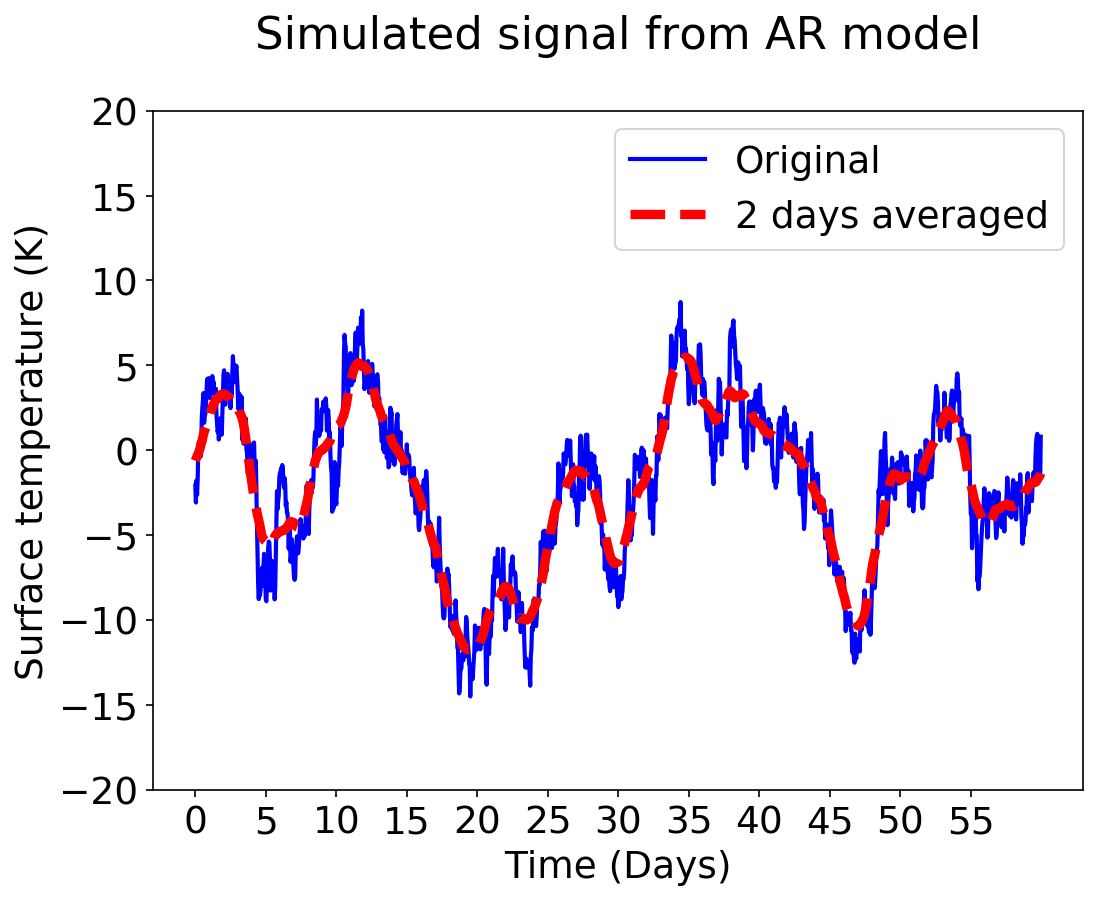}
\end{minipage}
\begin{minipage}{0.45\textwidth}
\textbf{\textbf{(b)}}\\
\includegraphics[height=0.9\textwidth,width=\textwidth]{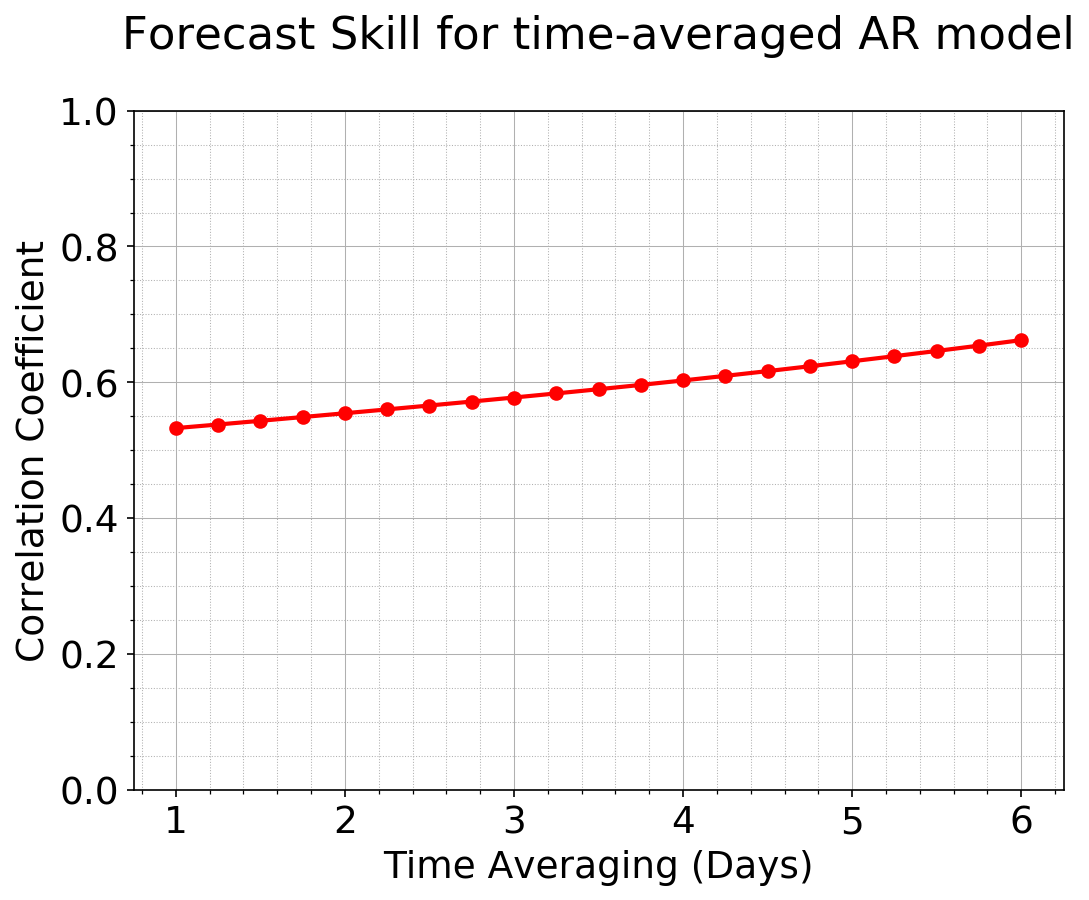}
\end{minipage}
\caption{Effects of time averaging on an AR(1) model. \textbf{(a)} An example time series of the simulated AR(1) process (blue solid line) and a smoothed version of it (red dashed line) obtained by time averaging with a 2-day window. \textbf{(b)} Forecast performance with different time averaging windows for a fixed lead time of 3 days for the AR(1) model.}
\label{OUfig}
\end{figure}

To better understand these results,
here we give a theoretical demonstration to show that 
if a time series is smoothed via averaging, 
it is not necessarily much more predictable. 
A theoretical illustration of this
can be seen using an auto-regressive (AR) model
(also called red noise, correlated noise, colored noise, or the Ornstein--Uhlenbeck process in the continuous-time version). The AR(1) model takes the form
\begin{equation}
    u^{n+1} = u^n - (\Delta t/T_d) u^n + \xi^n,
\end{equation}
where $u^n$ is the quantity of interest (e.g., precipitation
or surface temperature) at time step $n$,
$\Delta t$ is the time step,
$T_d$ is the decorrelation time,
and $\xi^n$ is a white noise process.
Such a model is often used to represent atmospheric and oceanic phenomena
\citep[e.g.,][]{d04,hm10b,ls20}.
An example time series is shown in Figure~\ref{OUfig}a. 
The example was created to be similar to a surface temperature
signal, and it uses a decorrelation time of $T_d = 3.8$ days.
Also shown in Figure~\ref{OUfig}a is a smoothed version of the AR(1) time series 
which was created by time averaging over a window of 2 days.
The time-averaged signal is much smoother than the original signal,
and one might therefore expect the time-averaged signal to be
much more predictable.

In Figure~\ref{OUfig}b, an experiment is implemented for testing the impacts of time averaging on the simulated time series from the AR process using 3 days as the lead time. One can see that, as the time averaging window increases, the forecast performance has only slight improvements, similar to what had been seen in the operational weather forecast data in Figure~\ref{fig:global-avg}.

Analytical formulas can be derived to better understand the
effects of time averaging on the AR model.
As one quantity of interest,
the decorrelation time provides a measure of predictability, 
and its value $\overline{T}_d$ for the time-averaged process can be calculated by 
\begin{equation}
    \overline{T}_d = \int_0^\infty ACF(\tau)d\tau
\end{equation}
where $ACF(\tau)$ is the autocorrelation function with time lag $\tau$. 
For the signal $u(t)$ from the original AR(1) model (in its continuous-time form as the Ornstein--Uhlenbeck process), the autocorrelation function is $ACF(\tau)=\exp(-\tau/T_d)$, and the decorrelation time is $T_d$.
For a time-averaged signal ($\bar{u}(t) =\frac{1}{T_w}\int_{t-T_w/2}^{t+T_w/2}u(s)ds$) 
averaged over a time window with length $T_w$, we find the decorrelation time to be
\begin{subequations}
\label{eqn:Td}
\begin{equation}
\overline{T}_d
= T_d \cdot\frac{(T_w/T_d)^2}{2[(T_w/T_d)-1+e^{-T_w/T_d}]},
\label{eqn:Td-exact}
\end{equation}
or, in approximate form,
\begin{equation}
\overline{T}_d
\approx T_d+\frac{1}{3}T_w.
\label{eqn:Td-approx}
\end{equation}
\end{subequations}
Details of the calculation of $\overline{T}_d$ are shown in the 
Supporting Information.

The analytical formula in (\ref{eqn:Td}) provides insight into
how the time averaging window length, $T_w$, affects the
decorrelation time, $\overline{T}_d$. When time averaging is performed
on the signal, how does the decorrelation time $\overline{T}_d$ of the smoothed signal compare with the original signal's decorrelation time, $T_d$?
As seen in (\ref{eqn:Td-exact}), $\overline{T}_d$ is proportional to
$T_d$, and the proportionality factor depends on the key quantity
$T_w/T_d$.  If $T_w/T_d$ is small---i.e., if the time-averaging window length is smaller than the original signal's decorrelation time---then time averaging has only a small impact on the decorrelation time and forecast performance.
For example, in Figure~\ref{OUfig}, where $T_d = 3.8$ days, a time averaging window of $T_w = 2$ days leads to a $\overline{T}_d$ that is only 18\% larger than $T_d$. This is a somewhat small change given what might be expected from how much smoother the time-averaged signal looks in Figure~\ref{OUfig}a. Even for larger values of $T_w$, which would create even smoother time series, the corresponding increase in $\overline{T}_d$ is somewhat small compared with what one might expect. For example, when $T_w=T_d$, so that the time-averaged signal is greatly smoothed, the corresponding decorrelation time is $\overline{T}_d\approx 1.36 T_d$, for a 36\% increase even in this largely smoothed case.  
Both of these examples are in line with the approximation,
shown in (\ref{eqn:Td-approx}), of $\overline{T}_d\approx T_d+(1/3)T_w$,
which suggests that the additional decorrelation time brought about
by time-averaging is only one-third as large as the time averaging
window duration $T_w$.
These slight changes in decorrelation time will then translate into
only slight changes in forecast performance.

\section{Conclusions}
\label{sec:concl}

An investigation was conducted to assess the effects of
time averaging and space averaging on forecast performance,
with the aim of doing so in a systematic way.
To do so, all parameters were held fixed (lead time, etc.)
except for changes in the averaging window.
Another aspect of the systematic setup is that
the time averages were conducted with a centered averaging window
from time $t-(T_w/2)$ to $t+(T_w/2)$.

One feature seen here is that the effects of time averaging and
space averaging can differ in different geographical locations.
Based on global maps and specific location examples, it was seen 
averaging effects for surface temperature are distinctly different 
in land regions, ocean regions, and coastlines.
For precipitation, on the other hand, averaging effects
appeared to be more randomly distributed over the globe and did not
show any particular geographic pattern.

Furthermore,
the results show evidence that time averaging actually does not provide a
substantial improvement in forecast performance, in comparison to spatial
averaging, at least over the range of scales of 1 to 6 days
and 100 to several thousand kilometers which were considered here.
Consistent evidence is seen
in forecast data from both operational weather forecasts and
simple stochastic models (either time series models or spatiotemporal
stochastic models) that can be studied analytically.
Taken together, this is evidence from a variety of models that 
time averaging may create a smoother time series that is not necessarily much more predictable.

One could speculate on the possible origin of
the conventional wisdom that time averaging should improve
forecast performance. 
It is possible that the conventional wisdom
arose in an early era of forecasting,
\begin{color}{trackcolor}
before the improvements that have been seen in recent decades
in methods for model initialization and data assimilation,
the use of ensembles, and improvements to the models themselves,
such as higher resolution
\citep[e.g.,][]{a01,btb15}.
We note that the impact of ensembles was not tested 
with the use of GFS data here, and it would be interesting to
investigate the effect of ensembles in the future;
as a result, the use of GFS data provides a testbed with
modern forecast techniques although where ensembles
are not part of those modern-era techniques.
If the initial conditions and forecasts were
more uncertain during earlier eras of forecasting, 
then a time average would possibly
have a larger impact on improving forecast performance.
\end{color}

It is also possible that the effects of time averaging depend on 
the time scales of the averaging. Here, lead times of one day to 
roughly one week
were investigated, along with time averaging windows of duration up to 
6 days. It would be interesting to investigate other time scales,
such as subseasonal to seasonal predictions or short-term climate
predictions
\citep[e.g.,][]{v07,retal15,mrr18}.
Nevertheless, one might expect the principles of the
theoretical AR model (Figure~\ref{OUfig}) to hold no matter the time scales.

\section*{Acknowledgements}
The authors thank Jeff Anderson for helpful discussion.
This research was performed using the compute resources and assistance of the University of Wisconsin-Madison Center For High Throughput Computing (CHTC) in the Department of Computer Sciences. The CHTC is supported by the University of Wisconsin-Madison, the Advanced Computing Initiative, the Wisconsin Alumni Research Foundation, the Wisconsin Institutes for Discovery, and the National Science Foundation, and is an active member of the Open Science Grid, which is supported by the National Science Foundation and the U.S. Department of Energy's (DOE) Office of Science. The research of SNS is partially supported by the US Office of Naval Research (ONR) grant N00014-19-1-2421.

\bibliographystyle{unsrtnat}

\end{document}